\documentclass[journal]{vgtc}                
\ifpdf
  \pdfoutput=1\relax                   
  \usepackage{graphicx}                
  \DeclareGraphicsExtensions{.pdf,.png,.jpg,.jpeg} 
\else
  \usepackage{graphicx}                
  \DeclareGraphicsExtensions{.eps}     
\fi%

\graphicspath{{figures/}{pictures/}{images/}{./}} 

\usepackage{microtype}                 
\PassOptionsToPackage{warn}{textcomp}  
\usepackage{textcomp}                  
\usepackage{mathptmx}                  
\usepackage{times}                     
\usepackage{cite}                      
\usepackage{tabu}                      
\usepackage{booktabs}                  

\usepackage[T1]{fontenc}
\usepackage{dfadobe}  
\usepackage[color=red]{todonotes}
\usepackage{algorithm2e}
\usepackage{amsmath}
\usepackage[percent]{overpic}
\usepackage{color, soul} 
\usepackage{graphicx}

\usepackage{hyperref}
\usepackage{xcolor}

\onlineid{0}

\vgtccategory{Research}
\vgtcpapertype{please specify}

\title{Visual Analysis of Multi-Parameter Distributions across Ensembles}

\author{Alexander Kumpf, Josef Stumpfegger, Patrick Fabian H\"artl, R\"udiger Westermann}
\authorfooter{
\item
All authors are with Technical University of Munich (TUM)\\
Alexander Kumpf: alexander.kumpf@tum.de\\
R\"udiger Westermann: westermann@tum.de

}

\shortauthortitle{Author \MakeLowercase{\textit{et al.}}Visual Analysis of Multi-Parameter Distributions across Ensembles}

\abstract{For an ensemble of data points in a multi-parameter space, we present a visual analytics technique to select a representative distribution of parameter values, and analyse how representative this distribution is in all ensemble members. A multi-parameter cluster in a representative ensemble member is visualized via a parallel coordinates plot, to provide initial distributions and let domain experts interactively select relevant parameters and value ranges. Since unions of value ranges select hyper-cubes in parameter space, data points in these unions are not necessarily contained in the cluster. By using a multi-parameter kD-tree to further refine the selected parameter ranges, in combination with a covariance analysis of refined sets of data points, a tight partition in multi-parameter space with reduced number of falsely selected points is obtained.  
To assess the representativeness of the selected multi-parameter distribution across the ensemble, a linked side-by-side view of per-member violin plots is provided. We propose modifications of violin plots to show multi-parameter distributions simultaneously, and investigate the visual design that effectively conveys (dis-)similarities in multi-parameter distributions. In a linked spatial view, users can analyse and compare the spatial distribution of selected points in different ensemble members via interval-based isosurface raycasting. In two real-world application cases we show how our approach is used to analyse the multi-parameter distributions across an ensemble of 3D fields. %
} 

\keywords{Distribution comparison, parallel coordinates brushing, violin plots}

\CCScatlist{ 
 \CCScat{K.6.1}{Management of Computing and Information Systems}%
{Project and People Management}{Life Cycle};
 \CCScat{K.7.m}{The Computing Profession}{Miscellaneous}{Ethics}
}

\teaser{	
	\includegraphics[width=\linewidth]{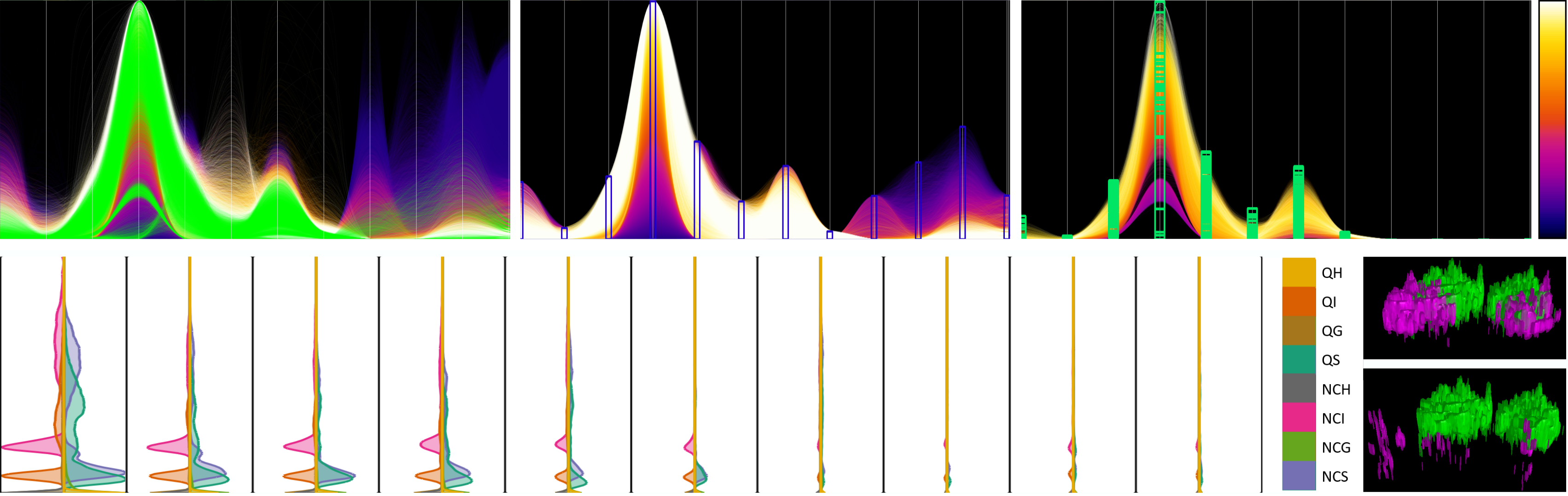}
	\centering
	\caption{Method overview: a) A parallel coordinate plot of 10 members of a multi-parameter cloud ensemble, each comprised of 250K data points with 12 physical parameters per point. Parameter space clustering in a representative member finds data points highlighted in green. b) A multi-parameter brush using the cluster's extreme values selects similar data points in all members. c) By using a kD-tree in parameter space, in combination with a principal component representation of the multi-parameter distributions of grouped data points, data points not following the clustered distribution are automatically discarded. d) A novel extension of violin plots for multi-parameter distributions enables to quickly reveal members with similar distributions. e) Via a linked 3D view, the spatial distribution of selected data points in different members can be compared and analyzed.
	}
	\label{fig:teaser}
}

\begin{document}

\firstsection{Introduction}
\maketitle
In many scientific fields like meteorology and computational fluid dynamics, numerical ensemble simulations are carried out with varying magnitudes of initial condition uncertainty, and by introducing uncertainty in the representation of certain physical processes. In such an ensemble, each simulation predicts possible states of physical quantities, aiming at a representative sampling of the physical phenomena that can occur. One of the major goals in ensemble visualization is to visually convey commonalities and differences between the ensemble members and, thus, to reveal the major trends and outliers in the simulation results.


Notably, while there is a considerable body of work related to the visual analysis of single-parameter ensembles~\cite{wang2018}, research dedicated to the analysis of multi-parameter ensembles, to the best of our knowledge, is rare. By a multi-parameter ensemble we mean an ensemble of fields where at each domain point a set of physical parameter values, the so called output parameters, is given. These fields are generated by using different model parameters, i.e., the input parameters of the simulation. For instance, in our current use case (Fig.~\ref{fig:teaser}) we work with an ensemble of 3D cloud simulations comprised of 96 members, each consisting of 250K data points with 12 different physical parameters.  

Previous works in multi-parameter data visualization, also termed multi-faceted, multi-field or multi-dimensional data visualization~\cite{KehrerHauser2013,liu2016visualizing}, have focused on visualizing the relationships between the parameters in a single data set, for instance, via parallel coordinates~\cite{inselberg1991parallel}, dimensionality reduction~\cite{JolliffePCA,  Bunte:2012:SNE:2207275.2207344}, 
or pair-wise scatterplot matrices~\cite{carr1987scatterplot}. In principle, such techniques can also be used to analyse the parameter variations across an ensemble, by either juxtaposition of single member visualizations or combined visualization of multiple members. In our use case, however, even for the visualization of a single member these techniques alone are not feasible due to visual clutter and over-plotting, and a combined visualization that can convey relationships between members becomes increasingly challenging.

An effective approach to reduce the number of entities that need to be considered when analyzing a multi-parameter data set is clustering in parameter space~\cite{kogan2006grouping,linsen2008surface, molchanov2018overcoming}. By finding groups of data points with similar multi-parameter values, cluster-based visual analysis using graphical abstractions for entire sets of data points can be utilized. 
However, when clustering is used with an ensemble---by clustering each ensemble member individually and comparing the results~\cite{Kumpf2019MultiParam}---differences in the number of clusters and their composition further complicate the analysis. In particular, the use of approximate cluster matchings to compare different clusterings can result in misleading decisions. 






To address these shortcomings, we propose an alternative visual analytics workflow for multi-parameter ensembles. This workflow allows the user to incorporate domain knowledge to select meaningful structures, and compare their occurrence across the ensemble.
Our approach builds upon the experiences we have gained when working with researchers from meteorology and computational fluid dynamics, where important structures are often described via specific combinations of multiple physical quantities. The workflow supports users in the selection of meaningful parameter combinations and value ranges of interest, and let them analyse whether similar structures occur in the ensemble and how representative they are.

\subsection*{Contribution}
We introduce a visual analytics solution to assess the occurrence of certain multi-parameter distributions in the members of a given simulation ensemble.   
To obtain an initial distribution that corresponds to a meaningful structure in the data, the system recommends a set of data points with similar multi-parameter values and shows them in a Parallel Coordinate Plot (PCP) (Fig.~\ref{fig:teaser}a). This set is obtained from a multi-parameter cluster in a representative ensemble member, by selecting data points that are withing the cluster's extreme values in each parameter (Fig.~\ref{fig:teaser}b). The system enables to assess whether the multi-parameter distribution over the selected data points is present in the other ensemble members, and where in the data domain the points contributing to this distribution are located. 


Since the proposed process selects a hyper-cube in the multi-dimensional parameter space, also regions in this space which do not contain any of the clustered data points are selected. When applying the selection to another member with data points in these regions, the multi-parameter distribution over the selected data points in this member is distorted.
In principle, this problem can be addressed by computing a tight hull of the clustered data points in the parameter space, and testing every data point in another ensemble member against this hull. However, since this approach is far too computationally complex in our scenario, we propose an alternative solution building upon an adaptive spatial subdivision scheme: Via a kD-tree over all clustered points in the parameter space, the selected intervals are automatically split into sub-intervals which more tightly enclose the clustered sub-regions (Fig.~\ref{fig:teaser}c). Via outlier removal and a covariance analysis of the locations of data points in each sub-region, the analysis can be focused on statistically representative sub-regions.   




To analyse the similarity of the multi-parameter distributions of selected data points across the ensemble, we propose a graphical depiction of the per-member distributions. This depiction is an extension of the classical violin plot ~\cite{ViolinPlots1998}, which we term Multi-Parameter Violin Plot (MPVP). An MPVP shows simultaneously the distributions of all parameters over a set of data points in a selected member. In the design of MPVPs, we have considered perceptual issues so that the user can quickly assess the major differences and similarities between the ensemble members. An MPVP plot showing all members simultaneously is linked to the PCP plot (Fig.~\ref{fig:teaser}d), and it is updated instantly when the user modifies the selection. 

A second linked spatial view shows the locations of the selected data points in the 3D domain. We follow the approach by Linsen et al.~\cite{linsen2008surface} to visualize a tight hull around the brushed data points via isosurface rendering. Since in our use cases the data points live on a voxel grid, surface extraction can be performed efficiently during GPU volume ray-casting. To compare different ensemble members to each other, rays are traced simultaneously against the hull of both the clustered data points and the data points in another member that is picked by the user (Fig.~\ref{fig:teaser}e). In this way, the user can quickly compare the spatial extent and mutual overlaps of regions showing a certain multi-parameter distribution. By picking the MPVP of a certain member, the spatial view is instantly updated to show the corresponding region.

The proposed visual analytics workflow builds upon the following specific contributions:
\begin{itemize}
    \item The combination of clustering and multi-parameter brushing to assess the occurrence of meaningful structures in multi-parameter ensembles. 
    \item An automatic refinement of multi-parameter selections using adaptive parameter-space partitioning and multivariate fitting.  
    \item Enhanced violin plots using multiple trace plots and difference-preserving coloring to simultaneously compare multiple parameter distributions. 
    \item A multi-parameter isosurface rendering technique, linked via a 3D spatial view, which indicates the parameter intervals that bound the cluster in the spatial domain.
\end{itemize}

To perform all operations interactively, we provide a flexible and scalable GPU rendering engine for parallel coordinate plots, including priority rendering, as well as linked MPVPs and 3D cluster boundary views. Fig.~\ref{fig:teaser} demonstrates the application of the proposed workflow to analyse a selected multi-parameter distribution in a numerical cloud ensemble, showing the specific visual encodings we propose. 

\section{Related work}

Our work is related to previous works in the fields of ensemble visualization and multi-parameter visualization. 

\paragraph{Ensemble visualization} Ensemble visualization is related to uncertainty visualization ~\cite{johnson2003next,bonneau2014overview}, yet it is assumed that the uncertainty is represented by a set of possible data occurrences rather than a stochastic uncertainty model. Previous works in ensemble visualization have addressed especially the question of how to visually convey the ensemble spread of certain physical fields, using either feature- or location-based approaches ~\cite{JoyFCEnsVis}. For a recent overview of ensemble visualization techniques let us refer to the survey by Wang et al.~\cite{wang2018}. Feature-based approaches, such as finding visual abstractions of the major trends in ensembles of line or surface features ~\cite{pang1997approaches,sanyal2010noodles,ContourBoxplotsWhitaker,FerstlEtAl2016EuroVis,DemirCentralTendency16}, are not subject of our study. 

Related to our approach are in particular approaches that represent the spread in scalar-valued ensembles via visual representations of statistical summaries ~\cite{love2005visualizing,PotterEtAl2010SummaryStat}.
Others model the variation of field variables by distributions, to reveal major trends and represent the ensemble in a compact way. 
Single-parameter ensembles are modeled via mixtures of probability density functions to compactly represent their spatial and temporal spread~\cite{liu2012gaussian,JarDemCVAEns,dutta2015distribution,wang2017statistical}.
Hazarika et al.~\cite{hazarika2017uncertainty} propose a copula-based framework for ensemble visualization, in particular to model the statistical dependencies between the scalar values at different locations. This approach was later extended to multi-parameter data ~\cite{codda2019}. 
Thompson et al.~\cite{thompson2011analysis} introduce so-called hixels, a meta-representation that encodes a histogram of scalar values in a certain spatial region.  
Multi-charts build upon a linearization of spatial locations and use side-by-side bar diagrams to compare the ensemble variability at different locations ~\cite{Demir2014MultiCharts}. H{\"o}llt et al.~\cite{HoelltViolin2013} and later He et al.~\cite{HE2020} use violin plots~\cite{ViolinPlots1998} to perform location-wise visualization of scalar time-series data. All these approaches visualize single-parameter ensembles, and cannot be extended in a straightforward way to multi-parameter ensembles. We extend these works by first selecting a set of data points with similar multi-parameter values in a reference member, and then determining and visualizing corresponding sets in all other ensemble members. 

\paragraph{Multi-parameter visualization}
At the transition between ensemble visualization and multi-parameter visualization, a number of approaches have been developed to investigate the relationships between multiple input parameters and a single output parameter of a simulation~\cite{BrucknerMoeller10resultDriven,bergner2013paraglide,torsney2011tuner,Sedlmair2014}. In contrast to our approach, which aims at a visual comparison of extremely large multi-parameter data sets, these approaches shed light on the sensitivity of simulation results to variations in the multi-parameter input configurations. Thus, at the core of these approaches is a visual analysis of the relationships between input configurations and simulation outputs, rather than the visualization of multi-parameter data sets. 

For the visualization of multi-parameter data, a number of different techniques exist, such as radar charts over pair-wise scatterplots and correlation heatmaps~\cite{KehrerHauser2013, liu2016visualizing}. Most of these techniques, however, even though effective for rather moderate amounts of data points, do not scale well in the number of points and are, thus, problematic in our scenario. Another technique to directly visualize multi-parameter data points is Parallel Coordinate Plots (PCPs). Popularized by Inselberg~\cite{Inselberg1985, inselberg1991parallel}, a multitude of methods have been proposed to improve the visibility of single data points in PCPs and reduce clutter. Many of these task-specific adaptations are surveyed in the summary of Heinrich and Weiskopf~\cite{heinrich2013state}. 
Johansson and Forsell~\cite{johansson2015evaluation} give an overview of user-centered evaluations of parallel coordinates, Dasgupta et al.~\cite{dasgupta2012conceptualizing} focus on the use of PCPs to convey uncertainty in the data. 
For an overview of techniques using multiple coordinated linked views including PCPs let us refer to the work by Roberts at al.~\cite{roberts2007state}.






In our proposed workflow, we let the user interactively select lines passing through certain parameter ranges via brushing in the PCP of all data points.
Ward~\cite{ward1994xmdvtool} introduce n-dimensional axis brushing, where the brushes are created manually using sliders.  
Fua et al.~\cite{fua1999hierarchical, fua1999navigating} enable viewing the data at multiple resolutions by means of hierarchical clustering. They propose structure-based brushing to select data lying on a wedge in the hierarchical space. Proximity-based coloring, zooming and fading of non-selected nodes in the hierarchy further improves the analysis.
Roberts et al.~\cite{SmartBrush2019} propose a brushing technique where a line-strip defines the centers of brushed intervals in each parameter dimension, and the width of selected intervals can be controlled interactively. Brush patterns can be translated vertically and across axes to reveal similar patterns in different parameter ranges. Multiple brushes can be combined and angular histogram glyphs show the value distribution and direction within a brush and axis. We make use of some of the proposed visualization options and integrate additional means into PCPs to automatically refine the set of brushed data points.  

Also related to our approach is multi-parameter clustering, one of the most effective methods in multi-parameter visualization to reduce the number of entities that need to be visualized. Clustering is either performed directly in the multi-dimensional parameter space ~\cite{kogan2006grouping,linsen2008surface,molchanov2018overcoming}, or via subspace clustering that finds clusters within a sub-set of all parameter dimensions~\cite{cheng1999entropy,nam2012tripadvisor,baumgartner2004subspace,ferdosi2011visualizing}. 
Non-axis-aligned sub-spaces have been clustered and visualized via dimensionality reduction techniques, i.e., by using projections into linear sub-spaces in which structures in the data points are maintained~\cite{kruskal1964multidimensional,williams2004steerable,morrison2002hybrid,maaten2008visualizing}. 
For visualizing multi-parameter clusters in the spatial domain, Linsen et al~\cite{linsen2008surface} propose isosurface rendering of the cluster boundary via scattered multi-parameter interpolation. 
Clustering in combination with PCPs has been used by Lex et al.~\cite{5613440}, by extending PCPs with colored matrices to analyse and compare the quality of cluster assignments using different algorithms. Long and Linsen~\cite{van2009multiclustertree} compute a hierarchy of high density clusters, which are analysed in parallel coordinates and linked views. The bundled parallel coordinates presented by Palmas et al.~\cite{palmas2014edge} use clustering to improve visual continuity and create links between axis aggregates. Matchmaker~\cite{5613440} introduces curved meta-links between axes. The use of clustering in an ensemble of multi-parameter data sets has been proposed by Kumpf et al.~\cite{Kumpf2019MultiParam}, by clustering each ensemble member individually and comparing the resulting sets of clusters. To do so, however, matchings between clusters in different ensemble members need to be established. Since the number and composition of clusters in each member can change significantly, this becomes prohibitively unfeasible in real-world applications.

\section{Method overview and Data}
\label{sec:overview}
\begin{figure}[h]
	\centering
	\begin{overpic}[width=.99\columnwidth, tics = 10, trim = 0 0 0 0, clip]
		{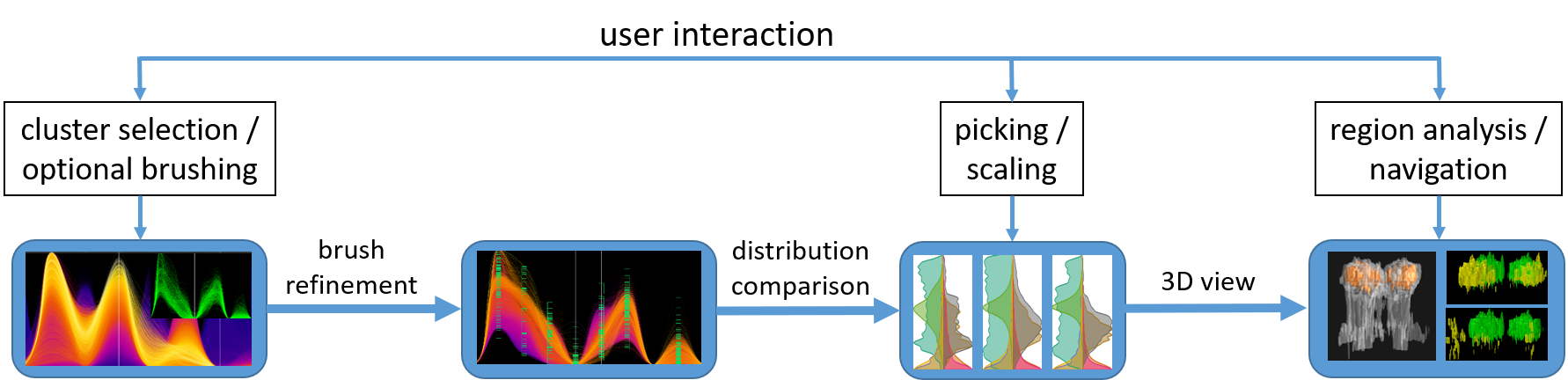}
	\end{overpic}
	\caption{Workflow overview: a) The user selects intervals in multi-parameter space by either selecting a computed cluster in that space or brushing manually in a PCP. b) Selected intervals are split adaptively using a regular spatial subdivision scheme and a non-axis-aligned refinement of sub-intervals. Refined intervals are applied to all ensemble members. c) MPVPs show instantly how representative a selected multi-parameter distribution is in all ensemble members, and how similar the distributions across the ensemble members are. d) A linked 3D spatial view shows the spatial locations of selected data points and can be used to compare the locations of data points in different ensemble members.}
	\label{fig:schematicWorkflow}
\end{figure}

Given an ensemble of multi-parameter data sets, we propose a visual analytics solution to interactively analyse the occurrence of a selected distribution of parameters in all ensemble member. Fig.~\ref{fig:schematicWorkflow} gives an overview of the workflow and the different visual analytics operations that are provided.


The user either selects the data in a pre-process, e.g., via clustering, so that certain multi-parameter structures are recommended by the clustering algorithm, or directly brushes parameter ranges of interest in the PCP interactively (Fig.~\ref{fig:schematicWorkflow}a). In the former case, the data in one representative ensemble member is clustered using any suitable clustering algorithm. For instance, HDBSCAN~\cite{campello2013density, HDBSCAN2017fast} after t-SNE projection of data points is used for the cloud data set. 

The representative ensemble member can be selected manually, for instance, by using domain knowledge. For the cloud ensemble, the experts deem the ensemble member that has been simulated with the parameter set closest to the mean of all parameter sets as a suitable representative. 
%
When working with ensemble data from the ECMWF Ensemble Prediction System (ENS; e.g., \cite{SwinbankEtAl2016}), an unperturbed control forecast (started from the ``best'' initial conditions) is used as representative member. This control run is available for all ensembles generated by ENS. 

A selected cluster in the representative ensemble member is shown in a PCP plot (see Fig.~\ref{fig:teaser}a), and converted to a \emph{multi-parameter brush} by using the extreme parameter values of all clustered data points. The user can further adapt the brush in an interactive and domain-specific way.
Following Roberts et al.~\cite{SmartBrush2019}, we use priority rendering of lines w.r.t. a selected value on one of the parameter axes. Therefore, all data points are sorted w.r.t. their distance to the selected parameter value, and blended on top of each other using adjustable opacity in the order of decreasing distance. By further mapping distances to colors, especially the perception of the density and spread of lines passing close to the selected parameter value can be enhanced significantly. 

In the next stage (Fig.~\ref{fig:schematicWorkflow}b), an automatic refinement of parameter intervals is computed, to filter out data points not belonging to the selected cluster, and the refined set of data points is visualized instantly (see Fig.~\ref{fig:teaser}c). 






Upon selecting suitable multi-parameter ranges, for each ensemble member a MPVP is generated to show the multi-parameter distribution of the selected data points (Fig.~\ref{fig:schematicWorkflow}c). Such a visualization is shown for the cloud ensemble in Fig.~\ref{fig:teaser}d. 
By picking an MPVP, the user can select an ensemble member and let it compare to the representative member in a 3D view (Fig.~\ref{fig:schematicWorkflow}d). Both the representative and the selected member are visualized via volumetric ray-casting, by rendering a hull with member-specific color around each set of selected data points (Fig.~\ref{fig:teaser}e). While the MPVP shows the distribution of parameters over a selected set of data point, the 3D spatial view shows where in the data domain these points are located and what shapes the corresponding structures have.   
Since the entire workflow is carried out on the GPU, all visualizations are instantly updated when the user performs certain interactions, like brushing in the PCP or picking an MPVP.

In the following, we discuss all stages of the proposed visual analytics workflow, and demonstrate its use for analyzing two real-world multi-parameter data sets. The first is an ensemble of $96$ numerical simulations of a growing thunderstorm cloud over a time span of 6 hours~\cite{wellmann2018Emulators}, simulated on a 700$\times$500$\times$35 Cartesian grid with anisotropic spacing. Of this data set ``Clouds''  we only consider the last time-step.
At each data point, 12 precipitation parameters---such as hail, water, and rain--- are given. 
Parameter vectors are first normalized over the whole ensemble, and data points with a norm less than $0.1$ are removed. Roughly 250k points per ensemble member remain of the initial set of points.
The second ensemble is an ECMWF weather forecast of tropical cyclone Karl initialized on 2016-09-22 00:00 UTC with 162h lead time, in an area from 30N to 80N and 50W to 30E for 30 vertical levels. This leads to roughly 120k grid locations in a $81\times51\times30$ grid.
Karl is comprised of 50 ensemble members and one control forecast simulated with best known initial parameters. 9 parameters like wind speed, temperature and precipitation parameters are considered.




\section{Multi-parameter brushing}
Brushing parameter ranges in a PCP to select data points corresponding to meaningful structures is challenging. 
Even though more sophisticated brushing techniques exist, such as area, lasso, or angular brushing ~\cite{raidou2015orientation}, we refrain from integrating such techniques into our workflow. This is in particular to relieve the user from specific assumptions about the relationships between different parameters, and let the selection be based solely on value ranges of the available physical quantities. Especially when working with experts from meteorology, we observed that meaningful structures are more or less exclusively defined via parameter intervals.

\subsection{Cluster-based brushing}
\label{sec:clbasedBrush}
Our strategy is to support the user in finding suitable initial parameter intervals. 
These intervals are further applied across the ensemble, by visualizing for each member the parameter distribution over the selected data points. This enables to analyse in which ensemble members a certain distribution is present, and how representative this distribution is for a structure in a certain member. Depending on the outcome of this comparative analysis, the multi-parameter intervals can be refined further via interactive brushing, and re-applied to the ensemble.   

In many practical applications, and even routinely in weather forecasting, important structures in a data set are often determined via clustering. Even though the results of clustering can sometimes be misleading, since they depend on intrinsic parameters of the clustering algorithm, extracted clusters often provide a good first guess about specific relations in the data. It is worth noting that in our application, where a cluster consists of a discrete set of data points, it is clear that the probability to find an identical cluster in some other ensemble member is very low. 

When an initial cluster in a multi-parameter data set is selected, the extreme values of each parameter over all data points in the cluster are used as a multi-dimensional brush in a PCP. Instantly, the selected data points are drawn over all other points in the PCP (Fig.~\ref{fig:teaser}a). 
Furthermore, we embed parameter histograms into the PCP, to convey information about the parameter distributions in the clustered data points, and compare them to the distributions of parameters in the entire representative member.

\begin{figure}[h]
	\centering
	\begin{overpic}[width=.99\columnwidth, tics = 10, trim = 0 0 0 0, clip]
		{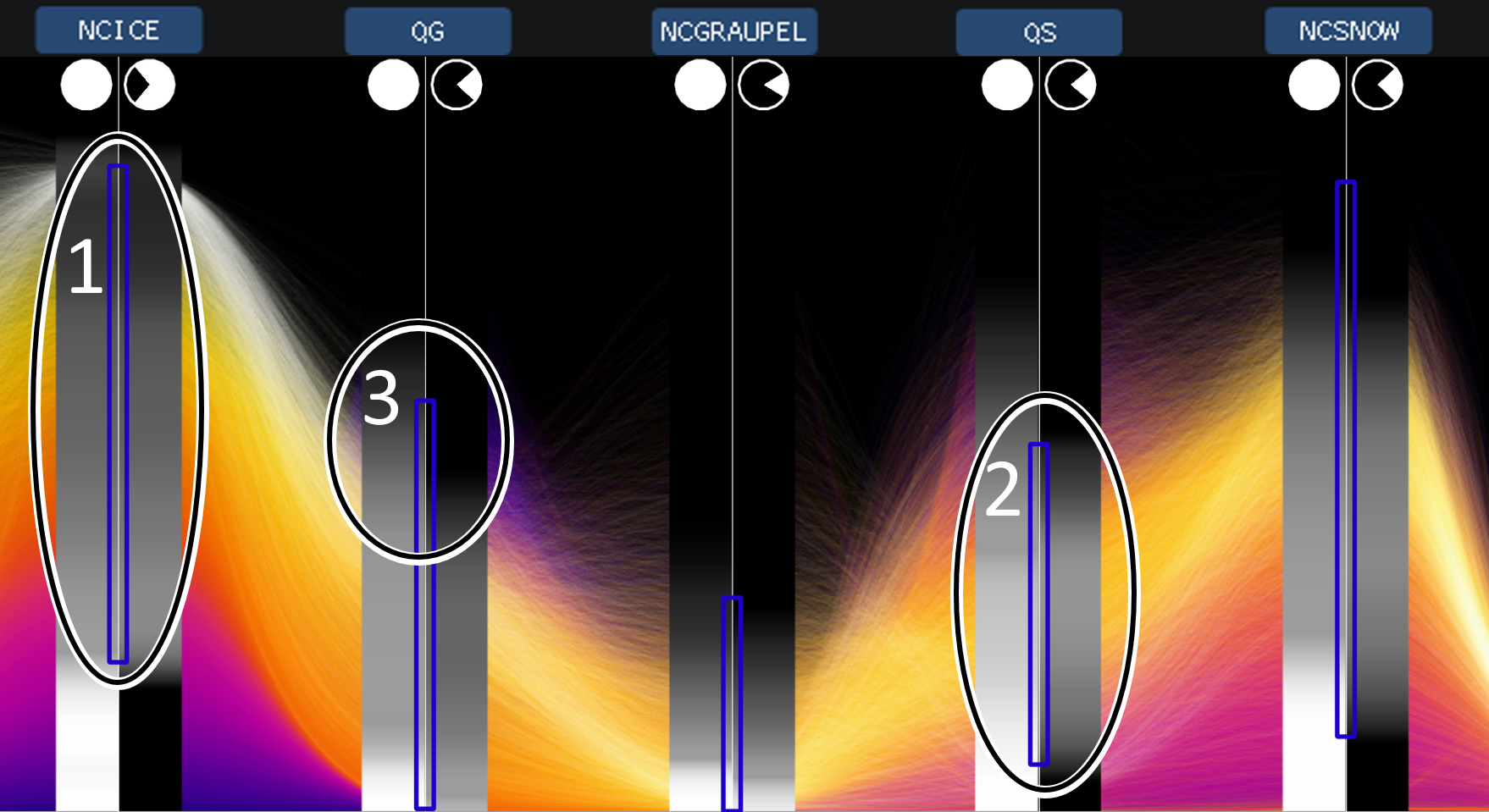}
	\end{overpic}
	\caption{
	%
	For each parameter, two histograms are drawn: One of the values in the data set (left) and one of the values in the selected cluster (right). Histograms indicate the representativeness of a parameter interval for a selected cluster, by showing the relation between the number of data points in a cluster and the overall number of data points in a certain parameter interval of the cluster. The latter is also shown by pie-charts above the cluster histograms.}
	\label{fig:priorRen_hist_cl_ds}
\end{figure}

As seen in Fig.~\ref{fig:priorRen_hist_cl_ds}, a side-by side view of parameter histograms for the entire and the clustered set of data points enables to quickly reveal two different aspects of the parameters: Firstly, they indicate how representative a parameter interval is for the selected cluster. For instance, the encircled interval 1 in the parameter range of ``number of ice particles'' is representative, since most data points where the number of ice particles is in that interval also belong to the cluster. Interval 2 in the parameter range of snow, on the other hand, is not representative, since for the majority of data points it cannot be decided whether they are in the cluster by only looking at the parameter snow. 
Note that few outliers in the cluster lead to an enlarged interval 3 for parameter graupel.
The percentage of data points in a certain parameter interval coming from a cluster is also shown via a pie-chart over the histogram for that parameter. Secondly, the cluster histograms reveal whether the distribution of parameter values in the cluster matches that of the data points, or the clustering algorithm has determined sub-structures in a certain parameter interval. 
In addition, via a bar-chart, the user is informed about the ratio of points in a cluster and points selected by the parameter intervals derived from the cluster.  
The ratio provides information about how representative the cluster is and how well the brush fits the cluster. The smaller this ratio is, the more points have been included that possibly do not belong to the clustered structure in the parameter space. 

\begin{figure}[h]
	\centering
	\begin{overpic}[width=.68\columnwidth, tics = 10, trim = 0 0 0 0, clip]
		{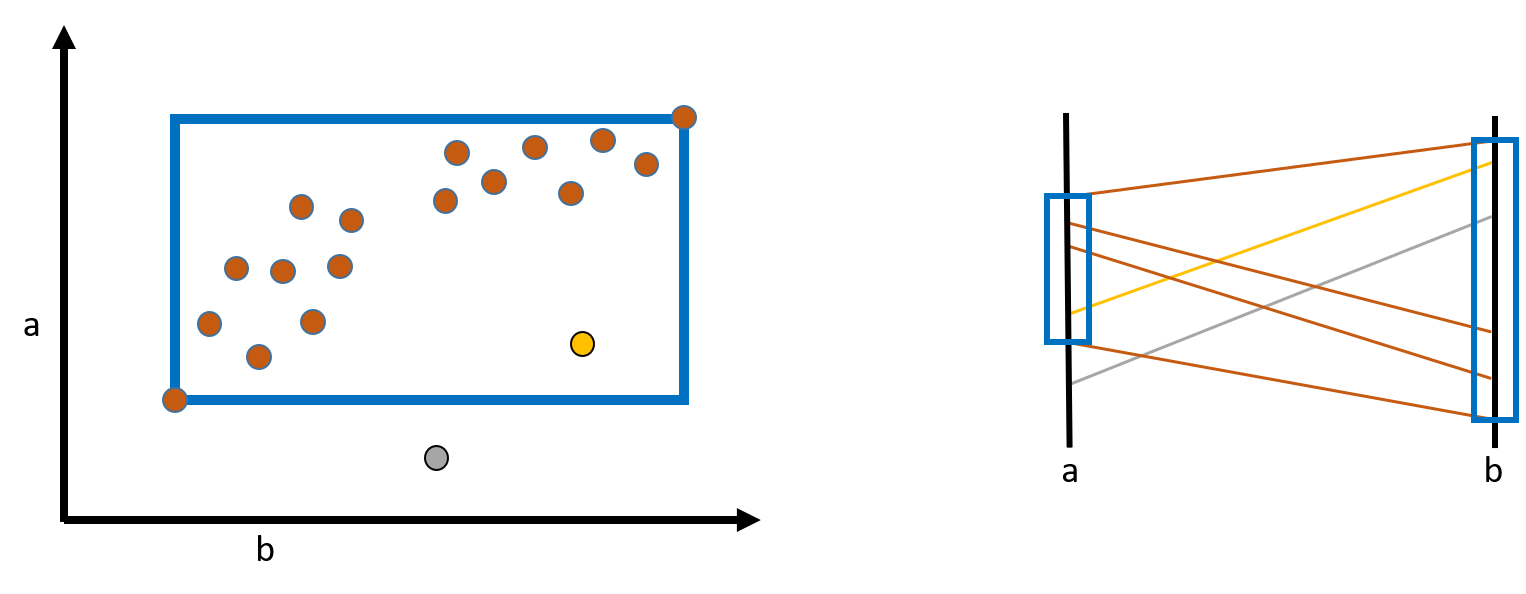}
	\end{overpic}
	\caption{Left: Two data points (yellow and grey) do not belong to a cluster in 2D parameter space (orange). The cluster's extreme values are indicated by the blue square. Right: The cluster is converted to a brush in a PCP (not all points are drawn). The data point colored yellow is falsely selected by the brush.}
	\label{fig:schematicBrushFromCluster}
\end{figure}

A problem that arises when brushing automatically w.r.t. to the interval ranges of a selected cluster is that the selected set of data points can include points that do not belong to that cluster. 
Cluster boundaries in parameter space, where the clustering algorithm operates, are usually not axes-aligned, but the intervals define an axis-aligned hyper-cube in the multi-dimensional parameter space. Hence, regions---and data points within them---are selected which are not covered by the cluster (Fig.~\ref{fig:schematicBrushFromCluster}).
Another problem arises if the selected parameter intervals are applied to select data points in another ensemble member. If this member has data points in a falsely selected region, the multi-parameter distribution over the selected data points in this member is distorted.

\begin{figure}[h]
	\centering
	\begin{overpic}[width=.99\columnwidth, tics = 10, trim = 0 0 0 0, clip]
		{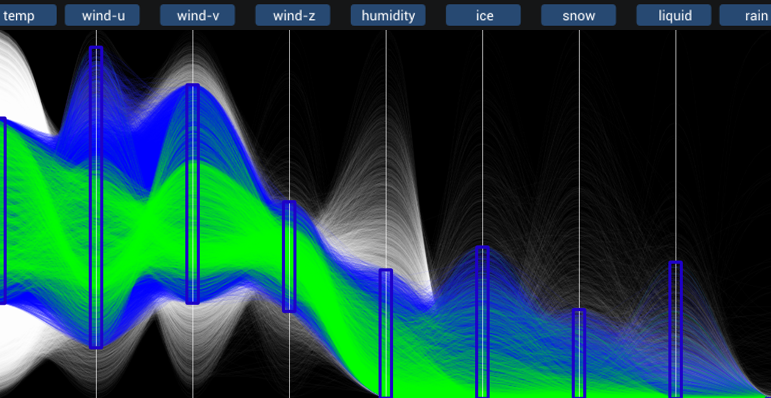}
	\end{overpic}
	%
	%
	\caption{Illustration of cluster-based brushing. (a) Green lines show a cluster of data points. Blue lines show the data points in that member that are selected by using the cluster's extreme values as a brush. 
	Points of the whole data set are rendered in white.}
	\label{fig:PCFilterComp}
\end{figure}

An example demonstrating the mismatch between clustered data points and data points selected by the cluster's extreme values is shown in Fig.~\ref{fig:PCFilterComp}, using the representative member in the ECMWF data set. The cluster (shown in green) is converted to a multi-parameter brush, which selects the data points rendered in blue. All remaining data points are rendered in white. The cluster is rendered last, so that blue lines shining through in the PCP hint to data points in the selected parameter intervals but not in the cluster.

\subsection{Brush refinement}
\label{sec:kdbrush}
The selection of data points using the extremal parameter values of a given cluster is sensitive to outliers and can include many more points than contained in the cluster. Furthermore, multiple disconnected sub-structures in parameter space can lead to large parameter ranges being selected. 
To achieve a tighter fit of the parameter intervals to the cluster, intervals are refined and re-applied to the data points. 
The effect of refinement is then also shown by pie-charts over the parameter histograms (Fig.~\ref{fig:priorRen_hist_cl_ds}), 
as the pie-chart of the brush and cluster become more similar.

\begin{figure}[h]
	\centering
	\begin{overpic}[width=.99\columnwidth, tics = 10, trim = 0 0 0 0, clip]
		{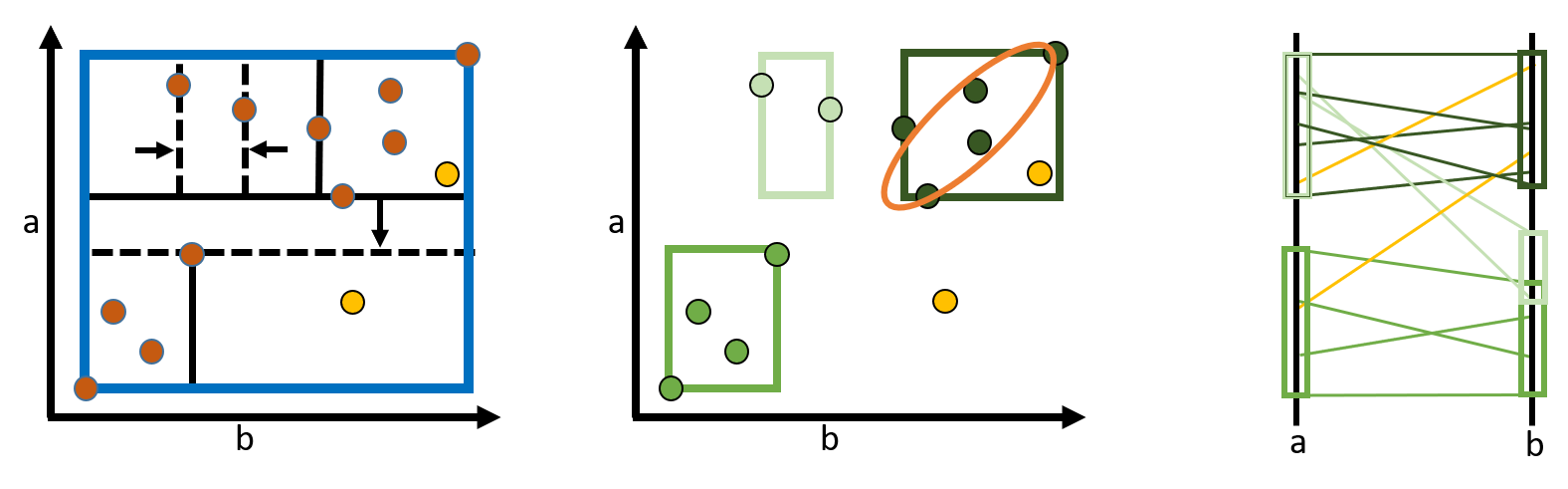}
	\end{overpic}
	\caption{Left: Parameter intervals derived from a cluster's (orange points) extreme values. First horizontal and then vertical kD-tree-based refinement is indicated by black lines. Dashed lines indicate narrowing of sub-intervals. Vertical refinement works on already narrowed sub-intervals. Yellow points do not belong to the cluster but lie in the derived brush. Middle: Resulting sub-intervals and covariance ellipse of selected point set. Right: Refined brush in PCP. Yellow points are not contained in either any sub-interval or covariance ellipse and can be filtered out.}
	\label{fig:schematicKDTree}
\end{figure}

For refinement, a kD-tree is constructed over the data points in the brushed D-dimensional parameter space. Thus, the initially selected hyper-cube is split into smaller cubes separating outliers and contiguous sub-structures. To build the kD-tree, we use the Surface Area Heuristic (SAH)~\cite{macdonald1990heuristics} to determine the locations of split hyper-planes along the parameter axes, in combination with a bounding interval-based approach to avoid the construction of very deep kD-trees. When splitting a certain parameter interval, the possible split locations are restricted to the parameter values of the data points in this interval. Therefore, all data points are first sorted w.r.t. the value of this parameter. Then, for all possible split locations the SAH measure $C$ is computed as 
\begin{equation}
    C=N_l \cdot L_l/L + N_r \cdot L_r/L. 
\end{equation}
Here, $N_l$ and $N_r$, respectively, are the number of points to the left and right of the current point (with the split point assigned to the sub-interval with more points or shorter length), $L$ is the length of the currently selected parameter interval, and $L_i$ and $L_r$, respectively, are the lengths of the left and right interval that are generated due to a split. Of all possibilities, the split point for which $C$ is minimized is selected, favoring large sub-intervals with low number of data points. Thus, outliers, contiguous subgroups and empty regions are effectively separated (Fig.~\ref{fig:schematicKDTree}). Due to alternating splitting along all parameter axes, the refinement considers the structure of data points in the multi-dimensional parameter space.

When using the SAH heuristic, however, a high refinement depth is required until empty regions are effectively separated. In Fig.~\ref{fig:schematicKDTree}, for instance, only when refining along the horizontal axis for the second time (after all parameter intervals have been refined once), the empty space between the isolated point at right and the cluster of data points to its left will be separated. To avoid this, after every split the interior boundary of the interval that does not contain the selected split point is refined, so that the resulting interval still bounds all points but is as small as possible (red boundary in Fig.~\ref{fig:schematicKDTree}). Thus, already after 1-2 refinements along every parameter axes a good fit is achieved.

\begin{figure}[h]
	\centering
	\begin{overpic}[width=.99\columnwidth, tics = 10, trim = 0 0 0 0, clip]
		{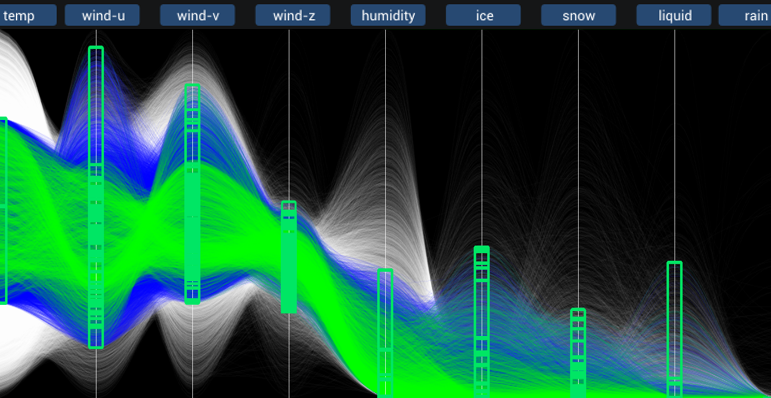}
	\end{overpic}
	\begin{overpic}[width=.99\columnwidth, tics = 10, trim = 0 0 0 0, clip]
		{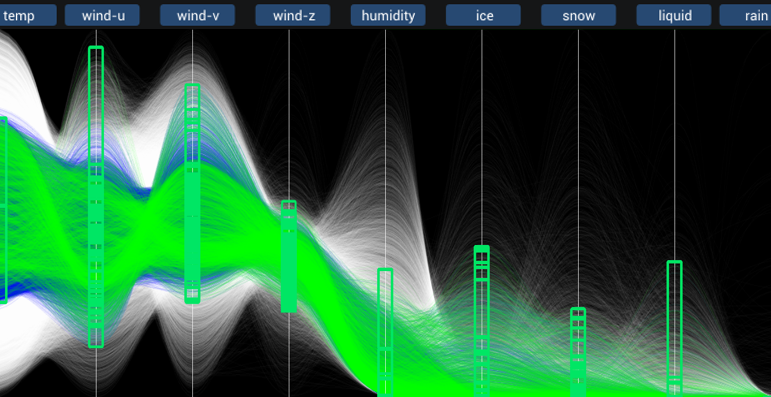}
	\end{overpic}
	\caption{Same as Fig~\ref{fig:PCFilterComp}, but intervals derived from the cluster's extreme values are first refined using the kD-tree-based interval refinement (top), followed by an additional distribution-based rejection of data points using a per-interval confidence region in principal component space (bottom).}
	\label{fig:ClusterRefine}
\end{figure}

Each refined hyper-cube is treated as an independent brush in parameter space. As shown in Fig.~\ref{fig:ClusterRefine}a, the proposed refinement can effectively prune outliers as well as empty space between them and the clustered structures. On the other hand, the distribution of data points in each hyper-cube can still be such that the derived interval bounds do not faithfully represent this distribution. 
For instance, the distribution of the data points in the upper right part in Fig.~\ref{fig:schematicKDTree} shows a clear diagonal preference, which is not well represented by axis-aligned intervals.   

To address this misalignment, we compute additional bounding representations that better fit the distribution of the data points in each hyper-cube. 
For this, we make use of a principal component analysis of the covariance matrix---containing the covariances between all parameter pairs of the data points---to determine a $\sigma$-confidence region in parameter space containing 68$\%$ of the data points. 
When a data point is tested for containment in a given multi-parameter brush, it is now tested for containment 
in the confidence ellipses representing the structure of points in these cubes. 
In this process we pay special attention to degenerated intervals and extremely small eigenvalues, and let corresponding dimensions solely be represented by the unit interval on the principal component axis.

As shown in Fig.~\ref{fig:ClusterRefine}b, the distribution-based refinement strategy prunes a number of additional data points that are not aligned with the majority of data points in a hyper-cube. This is also advantageous when applying the selected multi-parameter interval to other ensemble members, since it enables to consider the structure of data points in the parameter space rather than solely their locations.    

\section{Ensemble analysis}
Once a suitable multi-parameter brush has been obtained from the cluster's extreme values and further refined by using the kD-tree and covariance analysis, this brush is applied to all other  ensemble members. 
To reveal how representative the multi-parameter distribution of the selected points is for the ensemble, i.e., whether a significantly large set of data points with similar distributions exists in each member, we show the distributions for each member side-by-side. Juxtaposition of the single member visualizations can reduce cluttering, yet, on the other hand, the context between similar features in different members can get lost~\cite{gleicher2011visual}. While this is certainly a limitation of our approach, we try to reduce this limitation by a number of design decisions concerning the shape, color and ordering of the graphical elements used.  


\subsection{Multi-parameter violin plots}
\label{sec:violins}

The MPVPs we propose to visualize a multi-parameter distribution over a set of data points in a single member build upon density trace plots ~\cite{chambers1983graphical}, which were later modified by Hintze and Nelson~\cite{ViolinPlots1998} towards so-called violin plots. A violin plot is a vertical or horizontal density plot, with the probability density curve of a parameter on both sides of a vertical axis, optionally accompanied by a box plot of the parameter values in its interior. 
The book by Chambers~\cite{chambers2017graphical} gives a thorough overview of the different variants of this type of plot. H\"{o}llt et al.~\cite{HoelltEtAl2014} show the density curves of two different parameters on either side of the vertical axis to enable the comparison of two scalar value distributions. 
In a violin plot, multi-modal data distributions appear as multiple peaks in the density curve, and the distance between the curve from the axis provides information about the number of data points contributing to a certain parameter value.
The area under each density curve is either colored uniquely, or different colors are used in one plot to distinguish between multiple layers in the data.
Since a violin plot is simple and intuitive to understand, also many of them can be placed side-by-side without interference, making them useful for ensemble analysis.

\begin{figure}[h]
	\centering
	\begin{overpic}[width=1\columnwidth, tics = 10, trim = 0 0 0 0, clip]
		{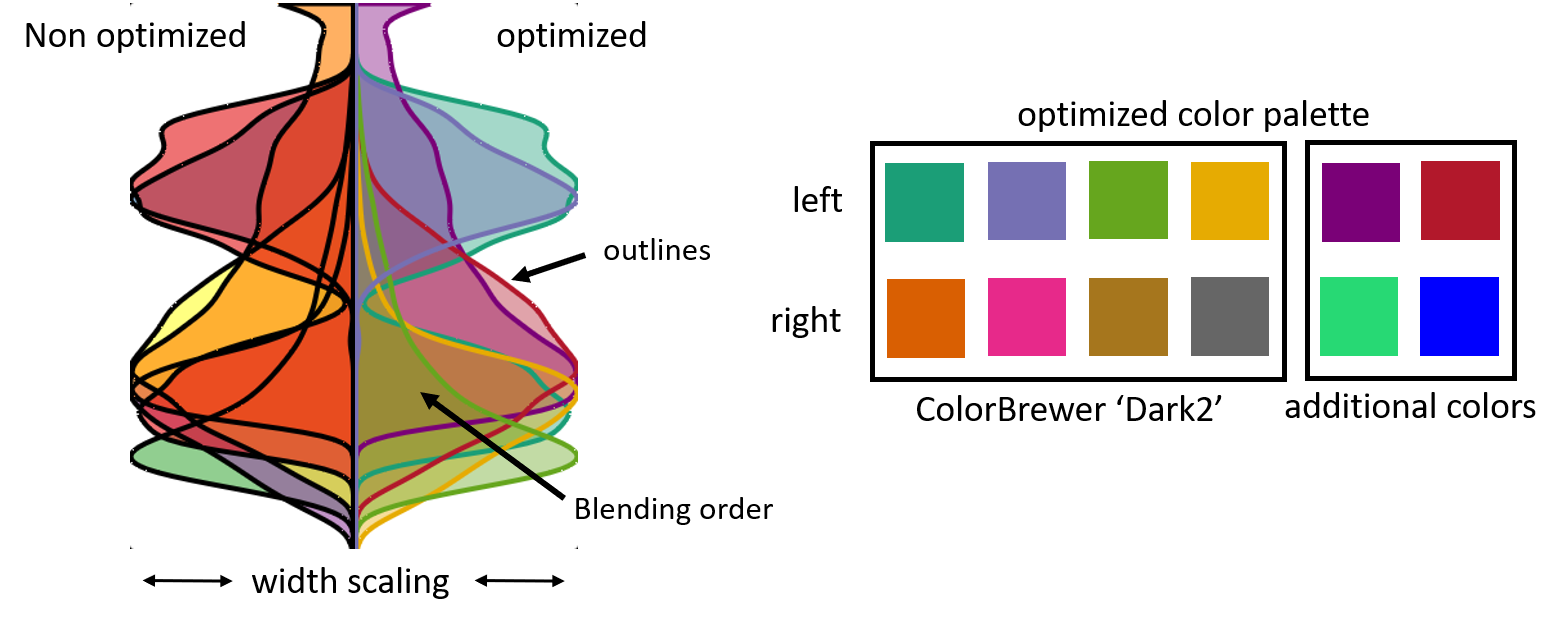}
	\end{overpic}
	\caption{Design parameters for MPVPs. Right of violin axis: Our favored design and colors. Left of violin axis: An alternative design that is less meaningful and requires more time for interpretation.}
	\label{fig:schematicViolin}
\end{figure}

The extension we propose to make a violin plot applicable to a multi-parameter distribution is to overlay multiple violin plots, each with its own colors on either side of the axis, in a single MPVP. This graphical depiction can also be used to show simultaneously the parameter curves of all ensemble members in one single plot (see the accompanied video). 
An illustration of the design parameters we have considered, with an ad-hoc selection on the left and our selected design parameters on the right, is shown in Fig.~\ref{fig:schematicViolin}. 

\begin{figure}[h]
	\centering
	\begin{overpic}[width=.8\columnwidth, tics = 10, trim = 0 0 0 0, clip]
		{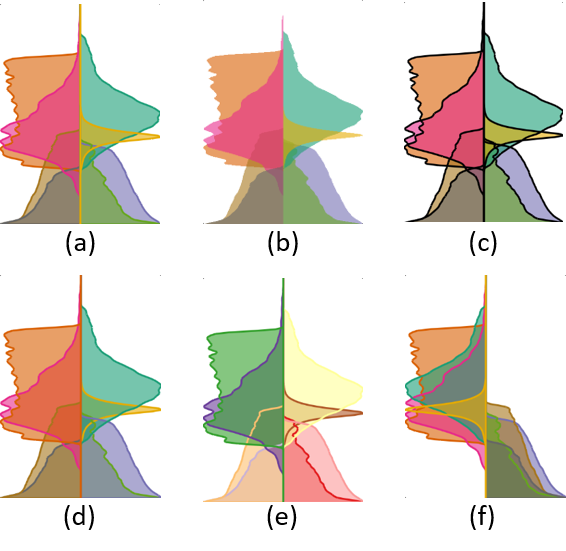}
	\end{overpic}
	\caption{Different visual designs of MPVPs. 
	a) Our proposed design with bold opaque outlines in the region color, as well as perception-aware color selection, left-right assignment of parameters, and drawing order. b) No outlines. c) Black outlines. a) Random drawing order and e) random colors of single violin plots.
	f) Random left-right assignment of parameter curves.}
	\label{fig:violin_renderOpts}
\end{figure}

The first design parameter we found important is the style and color of the outlines of the regions covered by the violins, i.e., the density curves. We evaluated three different possibilities (Fig.~\ref{fig:violin_renderOpts}): a) A bold outline in the color of the region. b) No outline. c) A black outline. Drawing no outline has two apparent drawbacks compared to the other options: Firstly, regions colored with a rather light color do not stand out against the background. Secondly, when regions overlap and the resulting color is similar to the color of the last rendered violin, the shape of the covered region can no longer be perceived clearly. Drawing a black outline can effectively reveal the shape of each region---also when blended with some other regions---yet since all regions have the same outline color it becomes difficult to distinguish between them where many contours appear. The first option (Fig.~\ref{fig:violin_renderOpts}a turned out to be most effective, standing out against the background and conveying the single plots in the overlap regions. 

When multiple violins overlap, their colors need to be blended. Overwriting colors by the color of the last drawn violin can occlude violins that are completely covered by this last one. One can also select a color for an overlap region that depends, for instance, on how many violins overlap in this region. This, however, makes it impossible to grasp the connection between an overlap region and the violins that contribute to it, since the violins' colors are not preserved in the color of the overlap region. In addition, this coloring results in abrupt color changes between the regions that occur due to different numbers of overlapping violins. Another issue that needs to be considered is that the visual context between multiple MPVPs should not be lost, meaning that overlap regions with the same set of contributing violins should be easily perceivable across a set of MPVPs shown side-by-side.    

To address these issues, we use $\alpha$-blending when drawing the single violins, and explicitly enforce a certain drawing order (Fig.~\ref{fig:violin_renderOpts}a). 
The violins are drawn in the order of decreasing area between the density curve and the violin axis, so that the chance to completely over-blend a single violin becomes lower, even though still possible.
The drawing order is established from the representative ensemble member and used for the visualization of all other members.
Violin plots are then blended atop of each other, i.e., we start with a white background  $C=(1,1,1)$ and then update this color via $C = \alpha \cdot C_V + (1-\alpha) \cdot C$ whenever drawing a violin with color $C_V$ and opacity $\alpha$. An $\alpha$ value of 0.4 is used to let overlaid colors shine through even when more than two layers are present. 
%
Fig.~\ref{fig:violin_renderOpts}d shows that some regions get lost when rendering and blending is performed in an arbitrary order.

To avoid hardly distinguishable blend colors, we follow the recommendation of Gama and Goncalves~\cite{gama2014studying}. As all complementary colors (being opposite on the color wheel) blend into a similar color, humans can better distinguish the blend colors of different pairs when these pairs are close on the color wheel. Since we have up to 12 parameters and, thus, need up to six colors on each side of the violin axis, we select two sets of 6 colors each. The 8 colors of the ColorBrewer palette Dark2 are sorted according to their hue, and split into two groups of 4 (right and left violins) such that the maximal distance between hues is minimal in each group (Fig.~\ref{fig:schematicViolin}). 
The additional 2 colors per side are chosen from the other side of the color wheel to avoid too many similar colors, yet with the trade-off of few complimentary colors.
Fig.~\ref{fig:schematicViolin}right and Fig.~\ref{fig:violin_renderOpts}a are both generated with these colors. In contrast, and as seen in Fig.~\ref{fig:schematicViolin}left and Fig.~\ref{fig:violin_renderOpts}e, when using an arbitrary color selection either very similar colors for different violins or pairs of complementary colors with the same blend color can be selected.

We further determine an assignment of ensemble parameters to the left or right of the violin axis that reduces the mutual overlaps between the single violin plots on either side.  
Therefore, a similarity matrix $D = \{d_{i,j}\}$ is computed, which represents the pair-wise overlaps of two parameter regions $i$ and $j$. 
Values $a_{ij}$ indicate the percentage of the region of $i$ that is covered by $j$. The minimum $d_{i,j} = \min{(a_{ij}, a_{ji})}$ defines the similarity between the violin plots of parameters $i$ and $j$. Large mutual overlap indicates overall similar shape of the plots, implying that the plots should be drawn on different sides of the violin axis. Small overlap, on the other hand, is not critical, since due to the drawing order the smaller plots are drawn atop of the larger ones. 
Iteratively, the pair of parameters with highest similarity value $d_{i,j}$ is determined, and these parameters are assigned to opposite axis sides. 
%
%
%
%
%
To decide which of the two parameters is placed left and which right, the violins already placed left $k_l$ and right $k_r$ are considered. In particular, the assignment is selected that minimizes the sum of the maximum similarities between the parameter pairs on the left and right side, i.e.,
\begin{equation*}
    \arg\min_{\rho\in\text{\{(r,l),(l,r)\})}}( \max_{k_{\rho[0]}}{d_{i,k_{\rho[0]}}} + \max_{k_{\rho[1]}}{d_{j,k_{\rho[1]}}} ).
\end{equation*}
Here, $r$ and $l$ indicate the right and left side, and $k_{\rho[{\cdot}]}$ indicates already assigned parameters on either side. As can be seen in Fig.~\ref{fig:schematicViolin}f, with no explicit assignment of parameter curves to the sides of the violin axis, large overlap regions can occur and absorb whole single violin plots.



\begin{figure}[h]
	\centering
	\begin{overpic}[width=.99\columnwidth, tics = 10, trim = 0 0 0 0, clip]
		{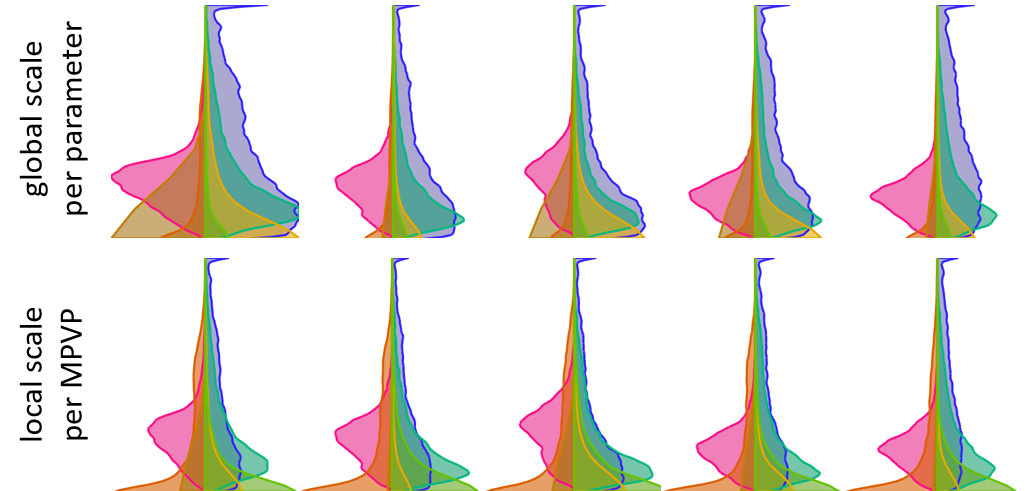}
	\end{overpic}
	\caption{MPVPs of 5 ensemble members. Top: For each parameter, the horizontal scale, i.e. the value at the left and right border, is set to the maximum value of the curves for this parameter over all ensemble members. Bottom: For each MPVP, the horizontal scale is set to the maximum value of all parameter curves of the corresponding ensemble.
	}
	\label{fig:violin_ScaleOpts}
\end{figure}

Another design decision is how to scale the single violin plots in the horizontal. Here, we provide two different scaling modes: The first one uses one scale for each parameter, and applies these scales to the respective parameters of every MPVP. For a particular parameter, the largest density value in any of the MPVPs is computed, and this value is set as the interval bound for that parameter in every MPVP  (Fig.~\ref{fig:violin_ScaleOpts} (top)). Even though this mode distorts the proportions between the parameters in one single MPVP, it is necessary when the total densities of parameters differ. It is in particular useful when showing multiple MPVPs side-by-side and comparing the parameter distributions across the ensemble. The second mode lets every single MPVP use it's own scale, which is applied to all parameter curves in this plot. In this way, the available drawing space in each plot can be used as good as possible, without sacrificing the relative proportions between the parameters (Fig.~\ref{fig:violin_ScaleOpts} (bottom)). This second mode is used when a fine-granular analysis of the multi-parameter distribution in a single member is desired.



\subsection{Linked spatial view}

\begin{figure}[h]
	\centering
	\begin{overpic}[width=.99\columnwidth, tics = 10, trim = 0 0 0 0, clip]
		{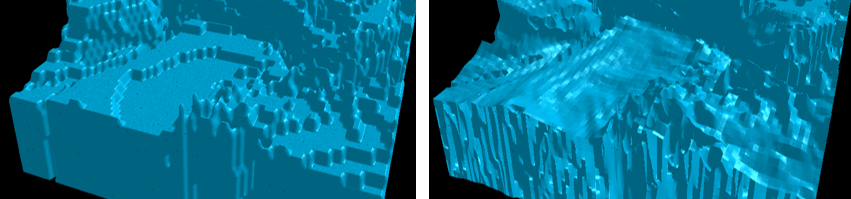}
	\end{overpic}
	\caption{a) Iso-surface in a binary field derived from a multi-parameter field of the representative member in Karl. b) Same iso-surface computed via smooth interpolation in each parameter field. 
	}
	\label{fig:rendercomparison}
\end{figure}

To further support the user in examining the spatial locations of selected data points, a 3D view is linked to both the PCP and MPVP chart.
Per default, the 3D view is synchronized with the PCP, so that any selection of a representative member or changes of the multi-parameter brush due to the selection of a cluster, refinement, or user interaction immediately triggers an update of the spatial view. In the spatial view, the locations of all data points in the selected member and the points in the refined cluster in that member are instantly rendered (Fig.~\ref{fig:rendercomparison}). The user can also select an MPVP, in which case the 3D view is linked to the ensemble member shown in this MPVP.
Then, instead of all data points in the representative member, the data points in the corresponding member that are selected via the refined multi-parameter brush are shown in green (e.g., top right in Fig.~\ref{fig:resultsKarlHalfpage}).   

Rendering the selected data points in 3D space is performed in two different ways (Fig. ~\ref{fig:rendercomparison}), depending on whether a high performance or high quality visualization is favoured. In either case, surfaces enclosing the data points are rendered semi-transparently, by using adjustable opacity values.   

\begin{figure}[h]
	\centering
	\begin{overpic}[width=.99\columnwidth, tics = 10, trim = 0 0 0 0, clip]
		{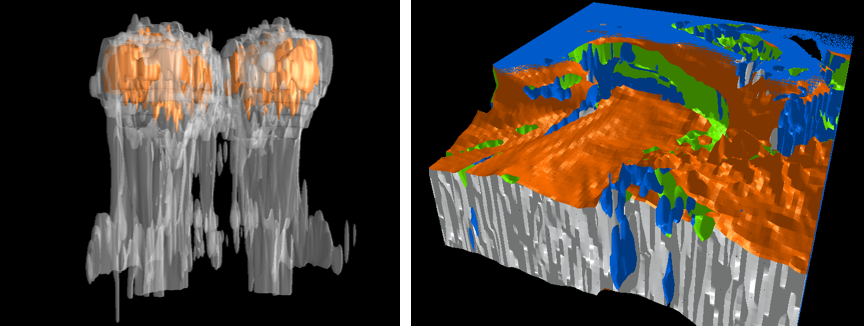}
	\end{overpic}
	\caption{a) Surfaces enclose all (grey) and brushed (orange) data points in the representative ensemble member of Clouds. b) Continuous multi-parameter surface  rendering in Karl shows which parameter intervals limit the surface structure in 3D space. Wind-speed in white, temperature in orange, humidity in green and precipitation quantities in blue.
	}
	\label{fig:3DRender}
\end{figure}

\paragraph{Binary representation} During interaction, for instance, if the user manually changes the multi-parameter brush, selected data points are converted into binary values and rendered via isosurface ray-casting in the resulting binary field. A multi-parameter value is set to 1, if all parameters are within the selected parameter intervals, otherwise the value is set to 0. In the resulting binary volume, ray-casting is used against the iso-contour to the value of 0.1 using tri-linear interpolation. Optionally, the binary value distribution can be smoothed using an adaptive Gaussian low-pass filter, which keeps values of 1 unchanged to avoid removing small isolated data points. Since the computation of a binary volume is extremely fast on the GPU, even for Clouds with a size of $700x60x500$, fully interactive frame rates are achieved. 
Multiple binary fields can be stored and visualized simultaneously. We use this option to show the locations of brushed data points in the representative member relative to locations of all data points in that member (Fig.~\ref{fig:3DRender}a), and to compare the locations of selected data points in different member (top right in Figs.~\ref{fig:resultsKarlHalfpage} and ~\ref{fig:resultsCloudHalfpage}).

\paragraph{Continuous field representation} To generate a high quality visualization, ray-casting is performed simultaneously in all single parameter fields. During ray-casting it is checked at every sample point whether all interpolated parameter values are within the selected parameter intervals. The transition into and out of these regions is detected by a change from "at least one parameter out" to "all parameters in", and vice versa. For shading, the gradient in the field of the parameter that was last going "in" (when entering the region) and first going "out" (when exiting the region) is used. Computationally, this method is far more complex than the first rendering option due to the significantly higher texture accuracy and number of texture look-up operations during ray-casting. On the other hand, it considers the smooth variation of physical parameters across the domain and leads to a continuous multi-field visualization~\cite{Fofonov2016}. Furthermore, it can effectively show which single-parameter brush bounds the rendered surface, by coloring surface points according to which parameter triggered the surface hit (Fig.~\ref{fig:3DRender}b).



\section{GPU implementation}
\label{sec:gpu}


To enable interactive brushing and applying the brush instantly to all ensemble members, we have implemented a high-performance implementation on the GPU. 
It is in particular tailored to efficiently handle large multi-parameter ensembles, to be able to instantly update PCPs, MPVPs and 3D views upon user interaction. 

The GPU implementation is index based, so that once the data points with assigned index reside in GPU memory, many operations require touching on far less memory compared to an explicit representation. 
In order to support efficient and parallel range queries for brushing, a boolean buffer containing the activation of each index is present in each drawable instance. If a brush update occurs, a compute shader processes all indices of the drawable instance in parallel and updates the activation of each index. Afterwards a second compute shader then uses the so computed activations to update the indexbuffer for rendering.
Furthermore, histogram-based similarity queries between different ensemble members or between members and a given histogram can be performed in a highly efficient way. 
All data points can be processed in parallel, and by using atomic memory operations on the GPU, parallel increment operations on the values in the GPU histogram buffer can be issued in an exclusive way.    



Our implementation (code published on \url{https://github.com/}\footnote{Link removed for blind review.}) via the Vulkan graphics API enables cross-platform support and high rendering performance. On our target architecture---a standard desktop PC equipped with an Intel Xeon E5-1650 v2 CPU with 6$\times$ 3.50 GHz and NVIDIA GeForce GTX 1070 graphics card with 8 GB VRAM---the performance is high enough to interactively analyse the set of ensemble members via brushing, MPVP visualization, and volume rendering, i.e., roughly 2.5 million data points with 12 parameters each in Clouds. Upon loading the data, it takes roughly 0.5 seconds to execute a normal brush operation (2s for a multivariate refined kD-brush) and render the result for 10 ensemble members simultaneously, including histograms, and 200ms for a single member. At the same speed, the user can animate through the ensemble. Volume rendering using the binary and continuous representation takes roughly 0.1 sec and 6 sec for the larger data set Clouds.




\section{Application analysis}
\label{sec:results}


\paragraph{Tropical cyclone Karl} Fig.~\ref{fig:PCFilterComp} shows in green the cluster that is initially selected in the representative ensemble member, mainly concentrating in the region in which the storm is present. Cyclons contain fast moving air masses coupled with high precipitation quantities, which motivates the search for cyclones in other ensemble members using the selected multi-parameter distributions.

\begin{figure*}
	\centering
	\begin{overpic}[width=.99\textwidth, tics = 10, trim = 0 0 0 0, clip]
		{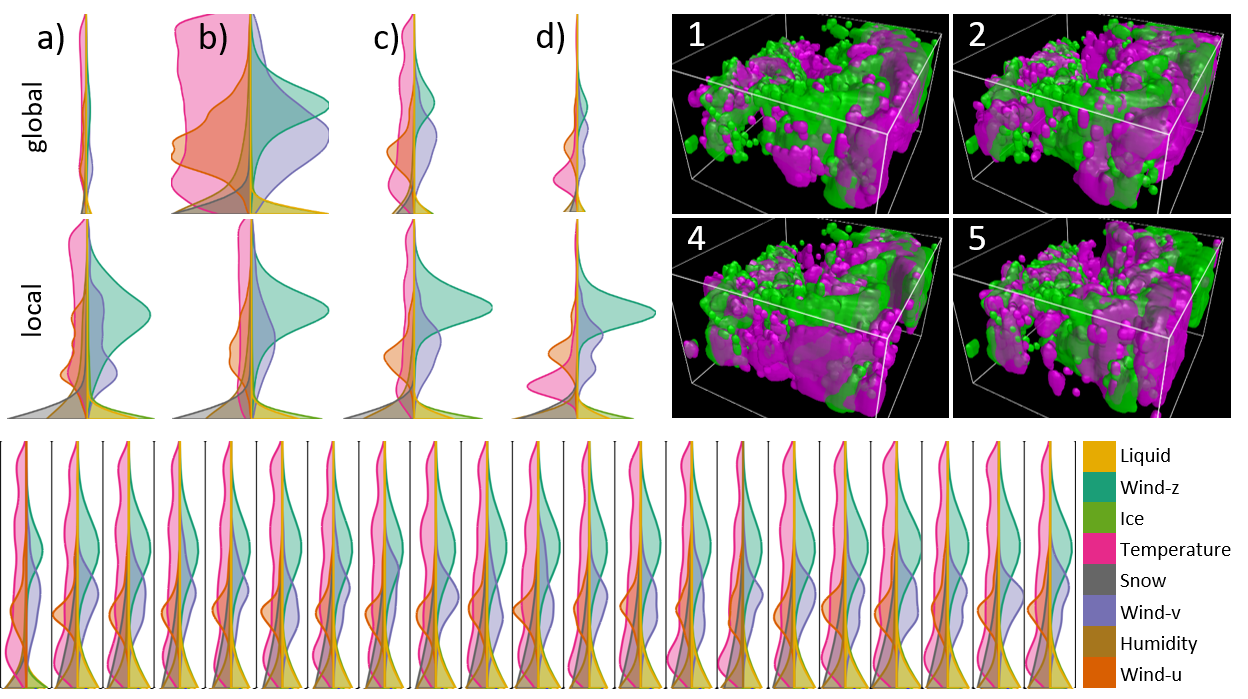}
	\end{overpic}
	\caption{Cyclon Karl. Top left: In the representative member, MPVPs with global and local scaling for data points a) in the cluster, and brushed via b) cluster extreme values, c) kD-tree-refined intervals, d) covariance-refined intervals. Top right: Locations of brushed points in the representative member (green) and members indicated by numbers (violet). Bottom: MPVPs with global width scaling for the representative and the first 20 ensemble members.}
	\label{fig:resultsKarlHalfpage}
\end{figure*}

Brushing with the cluster's extreme values per parameter selects roughly 15 times as many points in the representative member than in the cluster (blue lines in Fig.~\ref{fig:PCFilterComp}). By further refining the brush, the set of data points is significantly reduced, i.e., only 3 times as many points as in the cluster are brushed (shown in Fig.~\ref{fig:ClusterRefine}. This is also conveyed by Fig.~\ref{fig:resultsKarlHalfpage}a-d, where the MPVPs show the multi-parameter distribution over all clustered data points (a), over all data points that are selected by the initial brush in the clustered member (b), and over all points that remain in the parameter intervals refined via kD-tree (c) and covariance analysis (d). Covariance-based refinement with $\sigma$-confidence ellipse leads to a reduction of the number of selected points to a factor of 1.2, and especially the MPVP with local scale indicates only minor correspondence with the MPVP of the initially clustered points. Thus, it was decided to only refine via the kd-tree, and use corresponding MPVPs in the comparative visualization in Fig.~\ref{fig:resultsKarlHalfpage} (bottom). The comparison of 20 members using per-member MPVPs indicates very high representativeness of the selected multi-parameter distribution, both with respect to the number of selected points and the parameter distributions across these points. 

When visualizing the 3D locations of selected points in different members using the linked 3D spatial view (representative member and member 2 in Fig.~\ref{fig:resultsKarlHalfpage} (top, right)), a large structure corresponding to Karl is seen. Both ensemble members agree in the region in which selected data points are located and distinguish only in some smaller structures. 
We conclude that the multi-parameter distribution of the cluster can be used to select Karl in all ensemble members. A more fine-granular cluster or user-guided brushing would have to be used to further analyse occurring sub-structures.

\begin{figure*}
	\centering
	\begin{overpic}[width=.99\textwidth, tics = 10, trim = 0 0 0 0, clip]
		{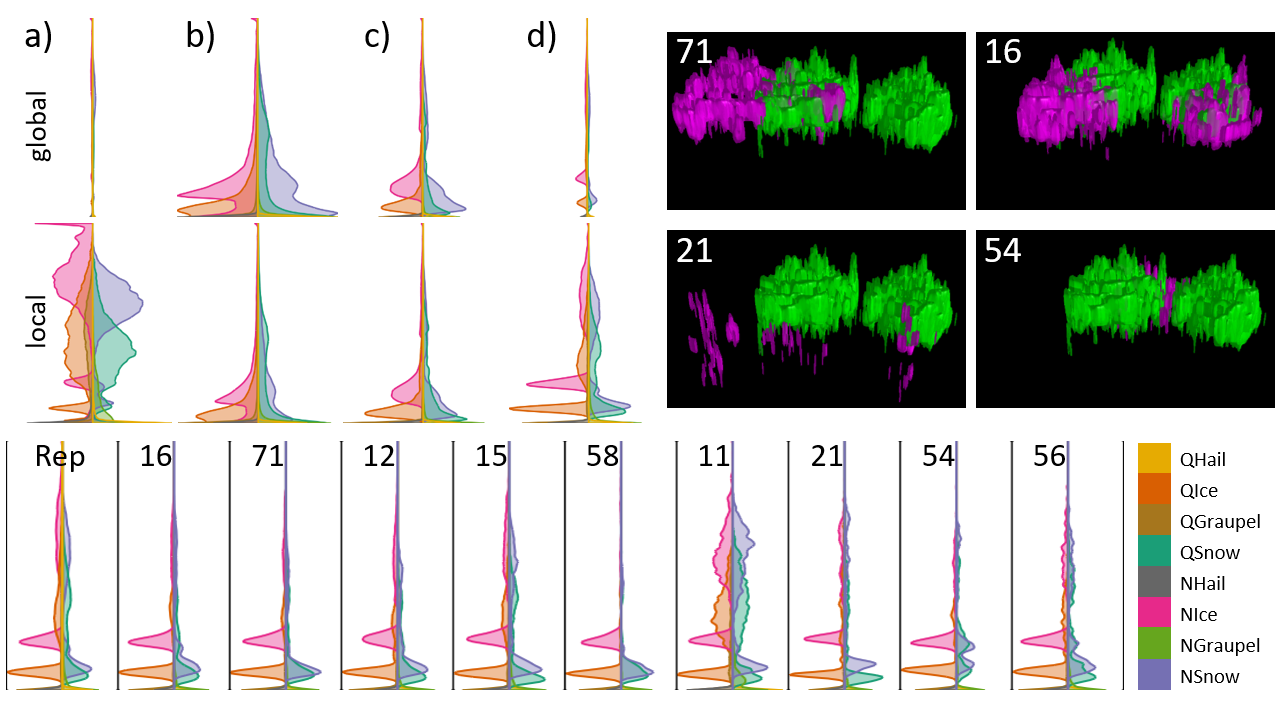}
	\end{overpic}
	\caption{Clouds. Top left: In the representative member, MPVPs with global and local scaling for data points a) in the cluster, and brushed via b) cluster extreme values, c) kD-tree-refined intervals, d) covariance-refined intervals. Top right: Locations of brushed points in the representative member (green) and members indicated by numbers (violet). Bottom: MPVPs with local width scaling for the same members as in Fig.~\ref{fig:teaser}.}
	\label{fig:resultsCloudHalfpage}
\end{figure*}

\paragraph{Cloud ensemble}
The initially selected cluster in Clouds shows two separate structures in the parameter NIce (Fig.~\ref{fig:teaser} and spatial view in Fig.~\ref{fig:3DRender}).
The same analysis as for Karl is performed. Fig.~\ref{fig:resultsCloudHalfpage}b-d show an effective reduction of the number of initially selected data points via kD-tree and covariance-based refinement. In particular, a brush solely based on the cluster's extreme values (Fig.~\ref{fig:teaser} (top, middle)), Fig.~\ref{fig:resultsCloudHalfpage}b)) cannot separate these structures and includes too many points. 
Covariance-based refinement leads to a multi-parameter distribution of selected points that is very similar to the distribution over the initially clustered data points (Fig.~\ref{fig:resultsCloudHalfpage}a), and it effectively separates the two cloud structures.

In contrast to Karl, the comparison of 20 members---chosen by k-Means on initial simulation parameters---using per-member MPVPs with global scale (Fig.~\ref{fig:teaser}) indicates far lower representativeness of the selected multi-parameter distribution, both with respect to the number of selected points and the parameter distributions across these points. It can immediately be seen that some members do not agree very well in the multi-parameter distribution of the selected data points. MPVPs with local scale (Fig.~\ref{fig:resultsCloudHalfpage} (bottom, left)) enable to asses the relative occurrences of parameters per-member distributions, yet a comparative analysis becomes meaningless. 
When visualizing the 3D locations of selected points in different members using the linked 3D spatial view (representative member and the most similar member 16 in Fig.~\ref{fig:resultsCloudHalfpage} (top,right)), one sees in particular very low agreement in the 3D locations of selected points. Both ensemble members agree in the shape of selected data points and distinguish only in some smaller selected structures. 

\section{Discussion and conclusion}
We have proposed visual analysis techniques and a workflow for analysing and comparing multi-parameter distributions in the members of an ensemble. These distributions are found by selecting data points in a representative member via brushing in the high dimensional parameter space, and applying the brush to all other members. The initial brush is automatically determined from a given cluster, and further refined to better match the single parameter distributions over the selected data points. To show simultaneously multiple parameter distributions over the data points in a single member, we have proposed an extension of violin plots (MPVPs) for multi-parameter data. As demonstrated for two real-world data sets, by using MPVPs the representativeness of a selected multi-parameter distribution for an ensemble is quickly conveyed, and outliers as well as other representative trends in the distributions are efficiently found. 
By a linked spatial view, the locations of selected points in different members can be visualized and compared to each other.  

In the future, we will consider alternatives to our proposed refinement strategy, which is based on the assumption of a normal distribution of points in each refined hyper-box. A purely stochastic refinement using Monte Carlo sampling might help to waive this assumption. Furthermore, we intend to apply the proposed workflow routinely at weather centres. Therefore, we see in particular the following two extensions that need to be further developed and integrated. Firstly, we need to incorporate automatic comparison measures for multi-parameter distributions, e.g., based on analytic approaches as proposed by Saikia and Weinkauf ~\cite{SaikiaWeinkaufTracking2017} for comparing histograms. Secondly, functionality to support the analysis of time-varying multi-parameter \emph{ensembles} is required. So far, our workflow has been demonstrated on a single time step, and it needs to be further investigated how the temporal changes of multi-parameter distributions over the ensemble can be assessed and compared in a meaningful way.  





\pagebreak



\textbf{Acknowledgements}\\
This research has been done within the subproject B5 of the Transregional Collaborative Research Center SFB/TRR 165 Waves to Weather funded by the German Research Foundation (DFG).



\bibliographystyle{abbrv-doi}

\bibliography{refs/BibPromo4}
\end{document}